\documentclass[]{aastex62} 
\usepackage{savesym}

\savesymbol{tablenum}
\usepackage[output-decimal-marker={.}, exponent-product=\cdot, per-mode=fraction]{siunitx}
\restoresymbol{SIX}{tablenum}

\usepackage{amsmath, amssymb, bm}
\usepackage{tabularx}
\usepackage{multirow}

\newcommand{\abs}[1]{\left| #1 \right|}
\DeclareSIUnit[]\rsun
{\text{R\ensuremath{_{\textup{sun}}}}}

\shorttitle{On the dependency between HSS velocities and CH areas}
\shortauthors{Hofmeister et al.}

\begin{document}

\title{On the dependency between the peak velocity of high-speed solar wind streams near Earth and the area of their solar source coronal holes}

\author{Stefan J. Hofmeister}
\affiliation{Institute of Physics, University of Graz, Austria}
\author{Astrid M. Veronig}
\affiliation{Institute of Physics, University of Graz, Austria}
\affiliation{Kanzelh\"ohe Observatory for Solar and Environmental Research, University of Graz, Austria}
\author{Stefaan Poedts}
\affiliation{Centre for mathematical Plasma Astrophysics (CmPA), Department of Mathematics, KU Leuven, Belgium}
\affiliation{Institute of Physics, University of Maria Curie-Sk{\l}odowska, Lublin, Poland}
\author{Evangelia Samara}
\affiliation{Centre for mathematical Plasma Astrophysics (CmPA), Department of Mathematics, KU Leuven, Belgium}
\affiliation{Solar-Terrestrial Centre of Excellence - SIDC, Royal Observatory of Belgium, Brussels, Belgium}
\author{Jasmina Magdalenic}
\affiliation{Solar-Terrestrial Centre of Excellence - SIDC, Royal Observatory of Belgium, Brussels, Belgium}

\begin{abstract}
The relationship between the peak velocities of high-speed solar wind streams near Earth and the areas of their solar source regions, i.e., coronal holes, is known since the 1970s, but still physically not well understood. We perform 3D MHD simulations using the EUHFORIA code to show that this empirical relationship forms during the propagation phase of high-speed streams from the Sun to Earth. For this purpose, we neglect the acceleration phase of high-speed streams, and project the areas of coronal holes to a sphere at \SI{0.1}{AU}. We then vary only the areas and latitudes of the coronal holes. The velocity, temperature, and density in the cross-section of the corresponding  high-speed streams at \SI{0.1}{AU} are set to constant, homogeneous values. Finally, we propagate the associated high-speed streams through the inner heliosphere using the EUHFORIA code. The simulated high-speed stream peak velocities at Earth reveal a linear dependence on the area of their source coronal holes. The slopes of the relationship decrease with increasing latitudes of the coronal holes, and the peak velocities saturate at a value of about \SI{730}{km.s^{-1}}, similar to the observations. These findings imply that the empirical relationship between the coronal hole areas and high-speed stream peak velocities does not describe the acceleration phase of high-speed streams, but is a result of the high-speed stream propagation from the Sun to Earth.   
\end{abstract}

\section{Introduction}
\label{sec:introduction}

\begin{figure}
    \centering
    \includegraphics[width=.5\textwidth]{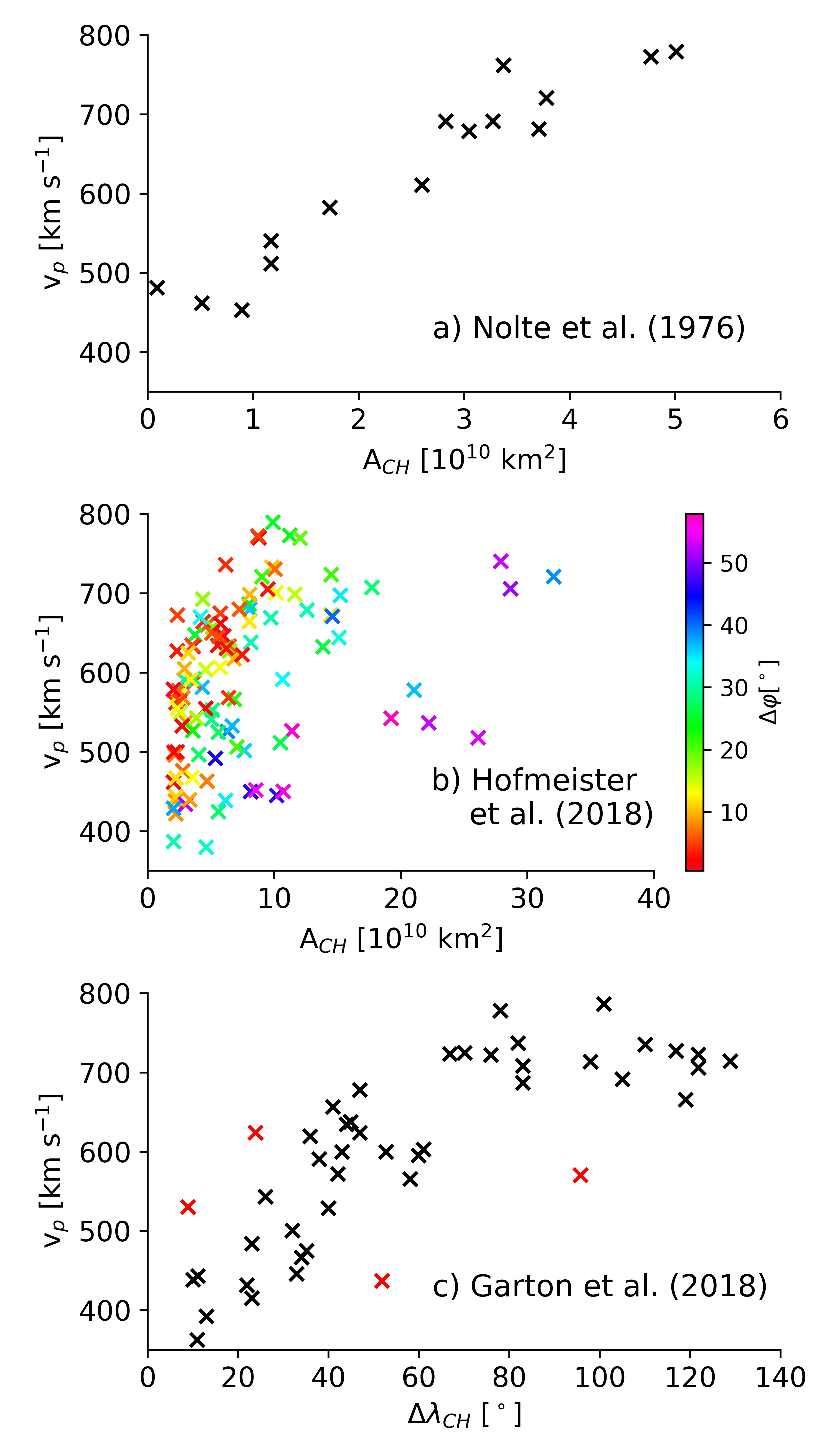}
    \caption{a) Peak velocities of three recurrent HSSs, which were observed in total 15 times by the Helios satellites at \SI{1}{AU} in 1973, versus the areas of their source coronal holes \citep[data from][]{nolte1976}. b) Peak velocities of 115 HSS observations by ACE, Stereo\-A and \-B during 2010 to 2017, versus the areas of their source coronal holes. The absolute value of the latitudinal separation angle between the satellites taking the HSS measurements and the center of mass of their source coronal holes are colour-coded \citep[data from][]{hofmeister2018}. c) Peak velocities of 47 HSS observations by ACE in 2016 and 2017 versus the longitudinal extensions of their source coronal holes; outliers are marked in red \citep[data from][]{garton2018}.}
    \label{fig:obsfacts}
\end{figure}

There are two well-known empirical relationships between the decisive properties of high-speed solar wind streams (HSS) and their solar source regions, i.e., coronal holes on the Sun: the relationship of the peak velocity of HSSs measured near Earth to (a) the area of coronal holes \citep{nolte1976}, and to (b) the inverse flux tube expansion factor of coronal holes \citep{wang1990}. While the relationship of the HSS peak velocity to the inverse flux tube expansion factor is usually interpreted in terms of the height at which energy is deposited to accelerate the solar plasma to form HSSs \citep{wang1991}, the relationship to the area of coronal holes is still physically not well understood. 

The linear relationship to the areas of coronal holes has first been reported by \citet[][Fig. \ref{fig:obsfacts} a]{nolte1976}, and thereafter confirmed by numerous studies \citep{robbins2006, vrsnak2007, abramenko2009, karachik2011, rotter2012, rotter2015, tokumaru2017, hofmeister2018, heinemann2018, heinemann2020}. The corresponding Pearson's correlation coefficients range from \numrange{0.40}{0.80}, depending on the composition of the datasets. The results of \citet{robbins2006}, who divided the Sun into three latitudinal regions of $\abs{\varphi} < \SI{30}{\degree}$, $\SI{30}{\degree} \leq \abs{\varphi} \le \SI{60}{\degree}$, $\abs{\varphi} > \SI{60}{\degree}$, already suggested that the slope of this linear relationship depends on the latitude of the source coronal hole. \citet{hofmeister2018} have investigated the dependence of this linear relationship on the position of the satellite taking the in-situ measurements relative to the coronal hole center, i.e., on their latitudinal separation angle. The authors found that the slope of the relationship between coronal hole areas and HSS peak velocities depends linearly on the latitudinal separation angle  (Fig. \ref{fig:obsfacts} b). The slope is steepest at separation angles close to \SI{0}{\degree}, i.e., above the center of the coronal hole, continuously declines with increasing separation angle, and basically turns to zero at a separation angle of \SI{60}{\degree}. 
Finally, \citet{garton2018} have shown that the peak velocity of HSSs are similarly well correlated to the longitudinal extension of coronal holes than to their areas, and that the peak velocities saturate at \SI{710}{km.s^{-1}} for coronal holes with a longitudinal extension $\SI[parse-numbers = false]{\gtrsim 67}{\degree}$ (Fig. \ref{fig:obsfacts} c).

In order to better understand the relationship between the peak velocities of HSSs near Earth and the areas of their source coronal holes, we employ a simplified magnetohydrodynamic (MHD) simulation setup for a parameter study. By projecting the areas of coronal holes to a sphere at \SI{0.1}{AU}, and propagating the resulting HSSs through the inner heliosphere using MHD simulations, we show for the first time that the relationship between the peak velocity of HSSs as measured near Earth and the area of their source coronal holes is a direct result of the evolution of HSSs during their propagation in the inner heliosphere. From this also follows that this empirical relationship is not related to the acceleration phase of HSS. In the Section \ref{sec:sim_setup}, we first describe the setup of our simulations. In Section \ref{sec:sim_results}, we  present and interpret the results of our simulations, and compare them to the observation. Finally, we formulate our conclusions in Section \ref{sec:conclusions}.

\section{Setup of the simulations} \label{sec:sim_setup}

\begin{figure*}
    \centering
    \includegraphics[width=\textwidth]{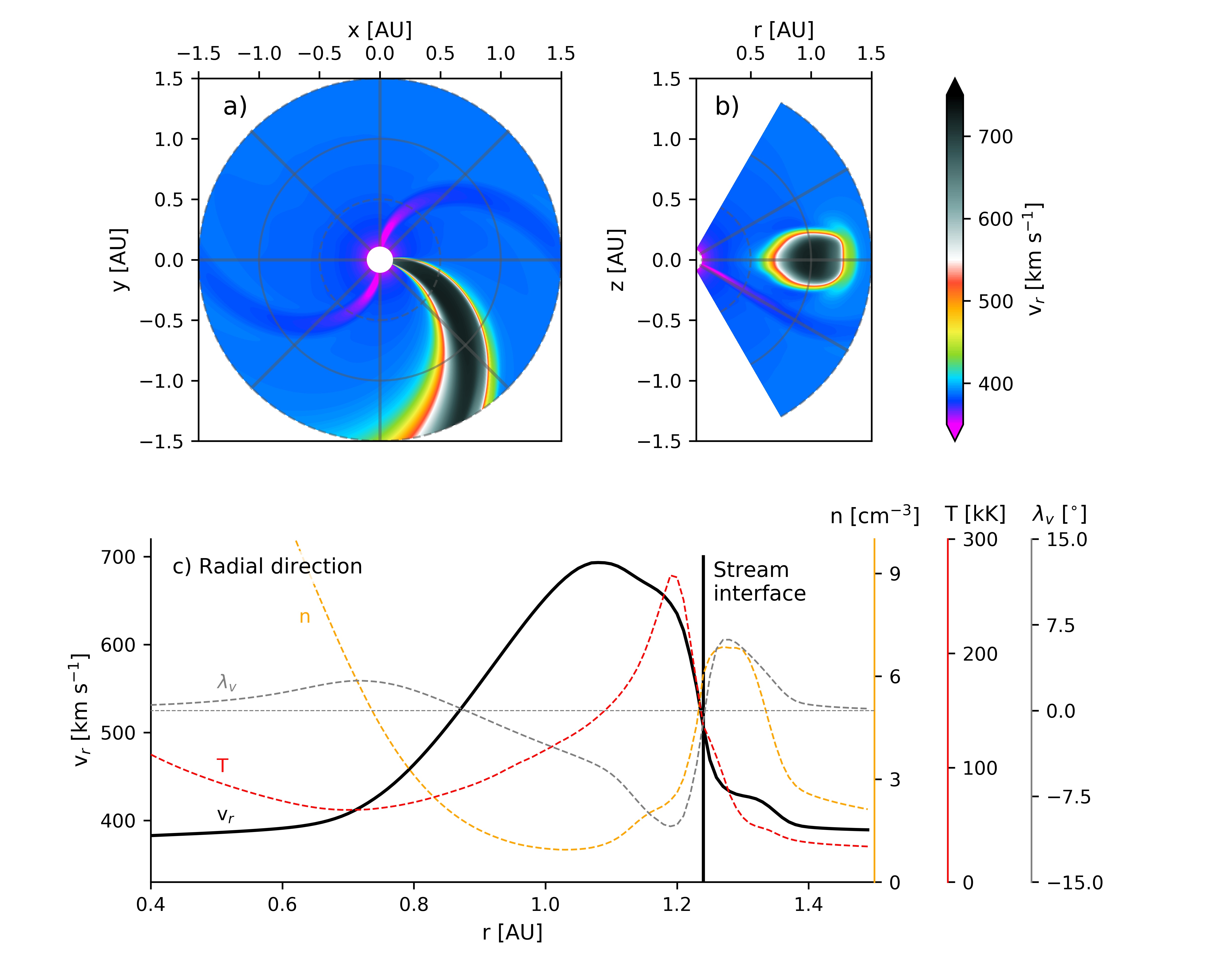}
    \caption{Simulation of a HSS in the inner heliosphere, derived by the EUHFORIA MHD code. a) and b) Snapshots of the radial velocity distribution in the solar equatorial and meridional plane. c) Snapshot of the properties of the HSS along a radial direction in the equatorial plane: radial velocity component (black), density (orange), temperature (red), and longitudinal angle of the flow velocity to the radial direction (grey). The grey dashed horizontal line corresponds to a longitudinal angle of \SI{0}{\degree}, i.e., a perfect radial flow. The black vertical line marks the position of the stream interface between the HSS and the preceding slow solar wind stream.}
    \label{fig:setup}
\end{figure*}

The European Heliospheric Forecasting Information Asset \citep[EUHFORIA, ][]{pomoell2018} was developed to model and predict the solar wind and the evolution of coronal mass ejections (CMEs) in the inner heliosphere for space weather forecasts. It consists of a simple, data-driven Wang-Sheeley-Arge (WSA) and Schatten-Current-Sheet (SCS) model of the solar corona which provides solar wind parameters at \SI{21.5}{\rsun}, and a 3D time-dependent MHD model of the inner heliosphere to propagate the solar wind  from \SI{21.5}{\rsun} to \SI{2}{AU}. CMEs can be injected in the model runs at \SI{21.5}{\rsun} using a cone or a spheromak model \citep{verbeke2019}. For furthers details, we refer to \citet{pomoell2018}.

In order to be independent of the not-well known acceleration phase of the solar wind from \SIrange{1}{21.5}{\rsun}, we assume a circular coronal hole at the base of the solar corona, skip the coronal model, and project the area of the coronal hole radially to a sphere at \SI{21.5}{\rsun}, resulting in the foot-print of the corresponding HSS at this sphere. We then set the velocity, temperature, and density of the HSS at this spherical shell to constant, homogeneous values, and propagate the HSS through the inner heliosphere using the heliospheric MHD model of EUHFORIA. Based on this idealized setup, we perform a parametric study by varying the size of the coronal hole and its latitudinal position while keeping all other parameters constant, in order to study their effect on the in-situ measured peak velocities at \SI{1}{AU} in the solar equatorial plane. 

At the spherical shell with radius \SI{21.5}{\rsun}, which corresponds to the inner boundary conditions of our simulations with EUHFORIA, we assume a bi-modal solar wind, and set the properties of the HSS and the surrounding slow solar wind for all simulations in the following way. For the HSSs, we set a uniform constant radial velocity of \SI{650}{km.s^{-1}}, density of \SI{150}{cm^{-3}}, and gas pressure  of \SI{3.3}{nPa}, and for the slow solar wind a velocity of \SI{350}{km.s^{-1}}, density of \SI{500}{cm^{-3}}, and gas pressure of \SI{3.3}{nPa}. These values result in typically observed velocities of about \SI{730}{km.s^{-1}} and densities of \SI{1.5}{cm^{-3}} for HSS plasma at \SI{1}{AU}, and \SI{390}{km.s^{-1}} and densities of \SI{5}{cm^{-3}} for slow solar wind plasma at \SI{1}{AU}. Further, we assume a lateral pressure balance at \SI{21.5}{\rsun} between the HSS and ambient slow solar wind plasma. Since the plasma $\beta$, i.e., the ratio of thermal to magnetic pressure, is $\ll 1$, i.e., the magnetic pressure dominates at these heights, we assume a uniform absolute mean magnetic flux density of \SI{217}{nT} both in the HSS and ambient slow solar wind region, which corresponds to a mean magnetic flux density of \SI{1}{G} at the solar surface. The magnetic field is assumed to be purely radial at the inner boundary, which also results in a radial direction of propagation of the HSSs. We added a heliospheric current sheet with a polarity inversion line specified by $ \varphi = \SI{-30}{\degree} \cdot  \cos{\lambda}$, where $\lambda$ and $\varphi$ are the solar longitude and latitude, respectively, and  set corresponding opposite magnetic field polarities at the northern and southern hemisphere in the slow solar wind. 

We then propagate these inner boundary conditions in time and space using the heliospheric part of the 3D MHD EUHFORIA code (Fig. \ref{fig:setup} a and b). We apply a fine spherical grid with a spatial resolution of \SI{0.006}{AU} in the radial direction from \SIrange{0.1}{1.5}{AU}, and \SI{1}{\degree} in the latitudinal and longitudinal direction. We relax the initial conditions over 17 days, which yields a steady, rotating solar wind solution. The solar wind properties are then extracted from the simulations using virtual in-situ spacecraft placed into the heliocentric equatorial plane at \SIrange{0.1}{1.5}{AU}, orbiting the Sun at a period of 365 days similar to Earth. In particular, we determine the peak radial plasma velocity $v_r$ of the HSSs at \SI{1}{AU} in the solar equatorial plane, and we derive the radial velocity $v_\text{SI}$ of the stream interface between the HSS and preceding slow solar wind plasma by its radial displacement with time.

Since the HSSs in our simulations propagate radially away from the Sun, it follows that the virtual spacecraft is in the latitudinal center of the HSS whenever the center of mass of the coronal hole is located in the solar equatorial plane. Shifting the center of mass of the coronal hole to higher solar latitudes also shifts the direction of propagation of the HSS to higher heliospheric latitudes; consequently, the virtual spacecraft located in the solar equatorial plane will take its measurements farther in the flanks of the HSS. Furthermore, since the plasma in HSSs propagates about radially away from the rotating Sun, the radial velocity component also approximates well the absolute value of the plasma velocity. The angle of propagation of the plasma to the radial direction does usually not exceed \SI{10}{\degree}. Finally, due to the very high magnetic Reynolds numbers in the solar wind, the magnetic field is frozen in the plasma, with the consequence that HSS plasma and the preceding slow solar wind plasma cannot mix. Instead, at the stream interface, a strong gradient in the solar wind properties is observable. The preceding slow solar wind plasma gets piled up by the faster HSS plasma, whereas the HSS plasma itself thermalizes and accumulates in the back of the stream interface (Fig. \ref{fig:setup} c). With increasing distance to the Sun, this stream interaction region develops into a forward and reverse shock pair. Since, at Earth distance, the magnetic field has an inclination of \SIrange{40}{60}{\degree} to the radial direction due to the rotation of the Sun, the slower preceding solar wind plasma is deflected by the faster impinging HSS plasma at the stream interface to the west, i.e., in direction of the solar rotation, whereas the radially propagating HSS plasma in its back is deflected to the east, i.e.,  along the direction of the Parker Spiral. Following \citet{borowsky2010}, we define the zero-crossing of the longitudinal deflection angle in this shearing region as the position of the stream interface.

\section{Results} \label{sec:sim_results}

\begin{figure}
    \centering
    \includegraphics[width=.5\textwidth]{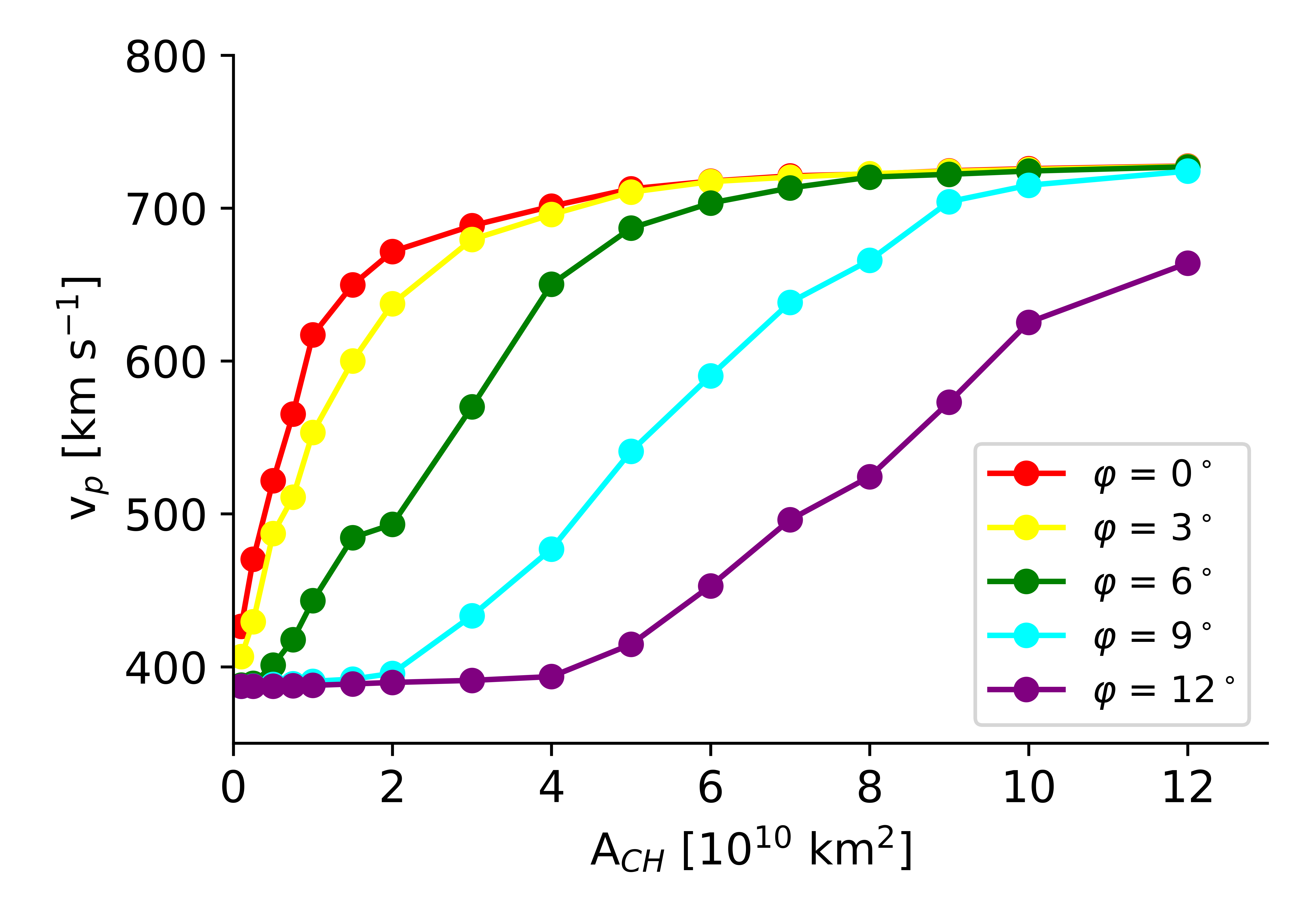}
    \caption{Peak solar wind velocities as measured by the virtual spacecraft versus the area of the coronal holes in the simulations. The virtual spacecraft was set to 1 AU in the solar equatorial plane. The latitudinal position of the coronal holes on the Sun is colour-coded. The velocity of the ambient slow solar wind at \SI{1}{AU} is \SI{390}{km/s}.}
    \label{fig:res_area}
\end{figure}

Since we want to compare our results with the empirical relationships derived from observations, we set up the study in a similar way. To reflect the evolution of an individual coronal hole and the resulting HSSs, we set a coronal hole to a given latitude and vary its size $A_\text{CH}$ in 16 non-uniform steps from \SIrange{1e9}{12e10}{km^2}. To reflect the variety of latitudes at which coronal holes appear, we vary the latitudes $\varphi$ of the center of masses of the coronal holes from \SIrange{0}{12}{\degree} at steps of \SI{3}{\degree}. 

The resulting peak solar wind velocities measured in the solar equatorial plane by our virtual spacecraft versus the coronal hole areas are plotted in Figure \ref{fig:res_area}, the latitudes of the coronal holes are colour-coded. These results match qualitatively the observations as described in Section \ref{sec:introduction}, and naturally show less spread than the observations due to the idealized setup. For each given coronal hole latitude, the HSS peak velocity increases linearly with the area of the coronal hole up to a given size, and then saturates at a constant value of \SI{730}{km.s^{-1}}. Thereby, the slopes of the linear relationships depend on the latitude of the coronal holes. For coronal holes located at the solar equator at $\varphi=\SI{0}{\degree}$, the peak velocities rise rapidly with increasing area up to the maximum velocity of \SI{730}{km.s^{-1}}, whereas the slopes are less steep for coronal holes located at higher latitudes. 

Since for all simulations we used the same properties for the HSSs at the inner boundary and only varied the latitudes and sizes of the source coronal holes, only the propagational evolution and the latitudinal direction of the HSSs can affect the peak velocities as measured by the virtual spacecraft at \SI{1}{AU} in the heliospheric equatorial plane. These findings strongly suggest that the relationship between the HSS peak velocities and coronal hole areas  is a direct result of the evolution of HSSs during their propagation in the inner heliosphere and the location of the spacecraft within the HSS. In the following sections, we further scrutinize the fundamental parameters of this relationship.

\subsection{Peak velocities of the HSSs in their latitudinal center} \label{subsec:maxv}

\begin{figure}
    \includegraphics[width=.5\textwidth]{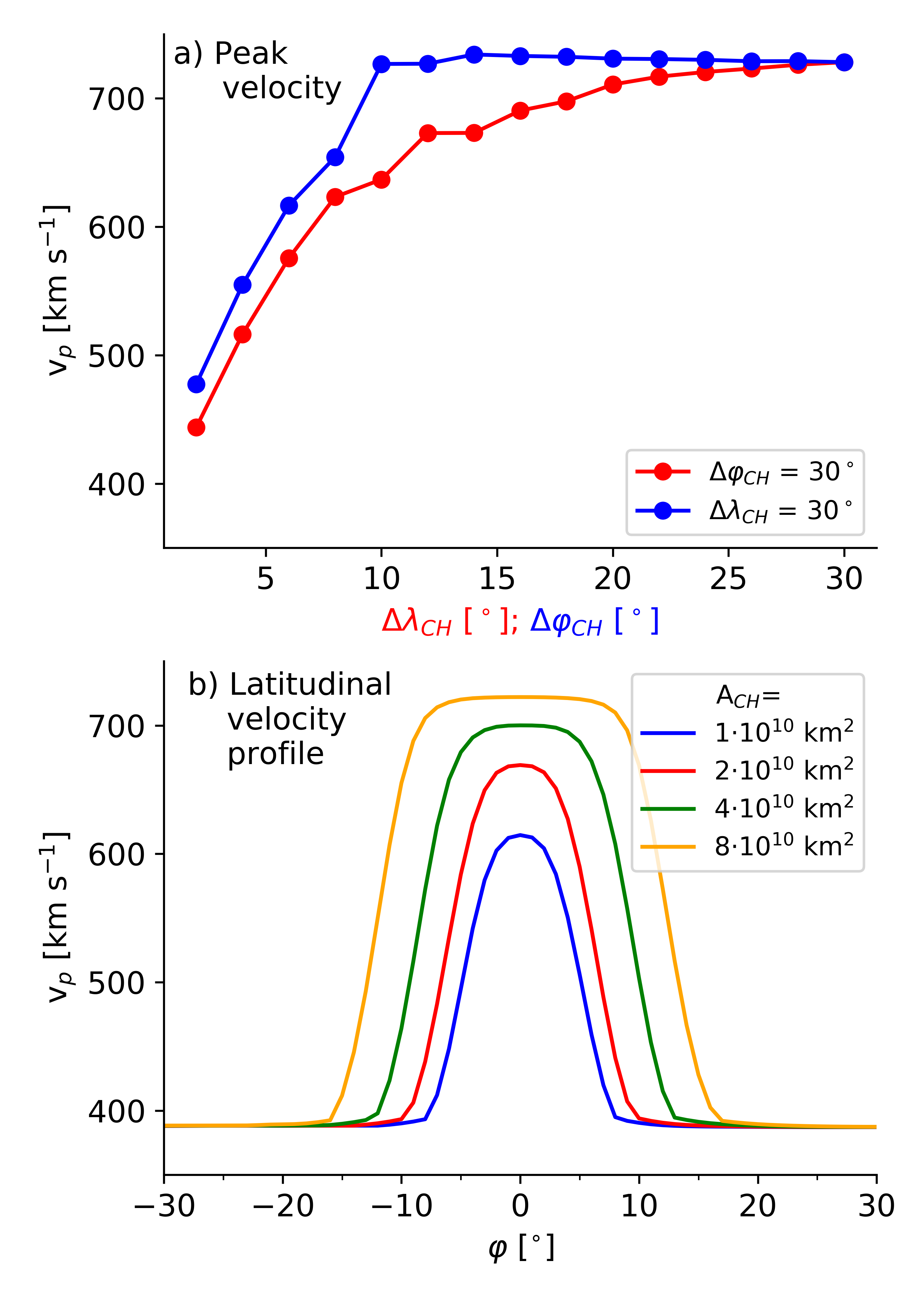}
    \caption{a) Peak velocities of the HSSs as measured by the virtual spacecraft in the solar equatorial plane at \SI{1}{AU} versus the width of rectangular coronal holes. In the first set (red), the latitudinal width of the rectangular coronal hole was fixed at \SI{30}{\degree} while the longitudinal width  was varied. In the second set (blue), the longitudinal width was fixed at \SI{30}{\degree} while the latitudinal width was varied.
    b) Latitudinal velocity profile of HSSs at a distance of \SI{1}{AU} to the Sun, for HSSs originating from circular coronal holes with areas of $1$ (blue), $2$ (red), $4$ (green), and $8\cdot$\SI{e10}{km^2} (yellow).}
    \label{fig:rectangular+lat}
\end{figure}

 First, we investigate the effect of the longitudinal width $\Delta\lambda_\text{CH}$ and latitudinal width $\Delta\varphi_\text{CH}$ of the coronal hole on the peak velocity of the HSSs in their center at 1 AU. To this aim, in this section, we set the center of masses of the coronal holes to the solar equator, and change its shape to rectangular. Since our virtual spacecraft is located in the solar equatorial plane and therefore in the latitudinal center of the HSS, it measures the maximum velocity of the overall stream. For the first set of simulations, we fix the latitudinal width at a large width of \SI{30}{\degree}, and vary the longitudinal width from \SIrange{2}{30}{\degree} (Fig. \ref{fig:rectangular+lat} a, red). For the second set of simulations, we fix the longitudinal width at \SI{30}{\degree}, and vary the latitudinal width from \SIrange{2}{30}{\degree} (Fig. \ref{fig:rectangular+lat} a, blue). 
 
 We find that increasing the longitudinal width of the coronal hole steadily increases the maximum velocity of the HSS over the whole range of the longitudinal widths, whereby the velocity increase with increasing longitudinal width slowly flattens out. In contrast, increasing the latitudinal width of the coronal hole steeply increases the maximum velocity of the HSS up to a width of about \SI{10}{\degree}, and then sharply saturates. Therefore, we may assume that up to latitudinal widths of \SI{10}{\degree}, erosion effects at the flanks of the HSSs affect the speeds in the center of the HSS. At larger width \SI{>10}{\degree}, the virtual spacecraft sees solar wind plasma which is undisturbed by its flanks.

Therefore, both the longitudinal and latitudinal width of the coronal hole affect the maximum velocity of the overall HSS. Increasing the longitudinal width increases the maximum speeds of HSSs in their center at \SI {1}{AU} up to large width of \SI{> 30}{\degree}, whereas the effect of increasing the latitudinal width on the peak velocities in the center of HSSs saturates at a latitudinal width of \SI{10}{\degree}.  

\subsection{Latitudinal velocity profile of HSSs}

Next, we investigate the latitudinal velocity profile of HSSs at 1 AU for coronal holes with areas of $1$, $2$, $4$, and $8\cdot$\SI{e10}{km^2}, using again circular shaped coronal holes (Fig. \ref{fig:rectangular+lat} b). For the large coronal hole of \SI{8e10}{km^2}, we find that the latitudinal velocity profile shows a broad plateau at \SI{730}{km.s^{-1}} around the center of the HSS, and that the velocities fall off steeply at the flanks to slow solar wind velocities. This plateau reflects the maximum velocity of the HSSs: further increasing the latitudinal width of the coronal hole, while keeping the longitudinal width constant, does not affect the maximum velocity of the HSS in its center, but only increases the width of the plateau. By decreasing the size of the coronal hole, we find that the plateau vanishes, and the maximum velocities in the center decrease. The gradual disappearance of the plateau in combination with the steep, but not abrupt decrease of the velocities across the flanks is a sign of erosion effects in the flanks of the HSSs. 

Note that even for the small coronal hole of \SI{1e10}{km^2}, the maximum velocity of the HSS, measured in its center, is still \SI{615}{km.s^{-1}}, i.e., changing the size of a coronal hole by a factor of 8 only reduces the maximum velocity of its associated HSS in its center at \SI{1}{AU} by \SI{115}{km.s^{-1}}. In contrast, changing the position of the satellite from the center to the flanks of the HSS strongly affects the peak velocities measured. By moving the latitudinal location of the virtual spacecraft by only \SI{10}{\degree} from the end of the plateau into the direction of the ambient slow solar wind, the peak velocities measured by the satellite change from large peak velocities close to the center of the HSS to small peak velocities close to that of the ambient slow solar wind. Therefore, the peak velocities measured are mostly affected by the relative location of the satellite within the HSSs. This relative location, again, is dependent on the latitudinal separation angle of the satellite to the center of the coronal hole, and on the area of the coronal hole.

\subsection{Radial evolution of HSSs}

\begin{figure*}[t]
    \centering
   \includegraphics[width=\textwidth]{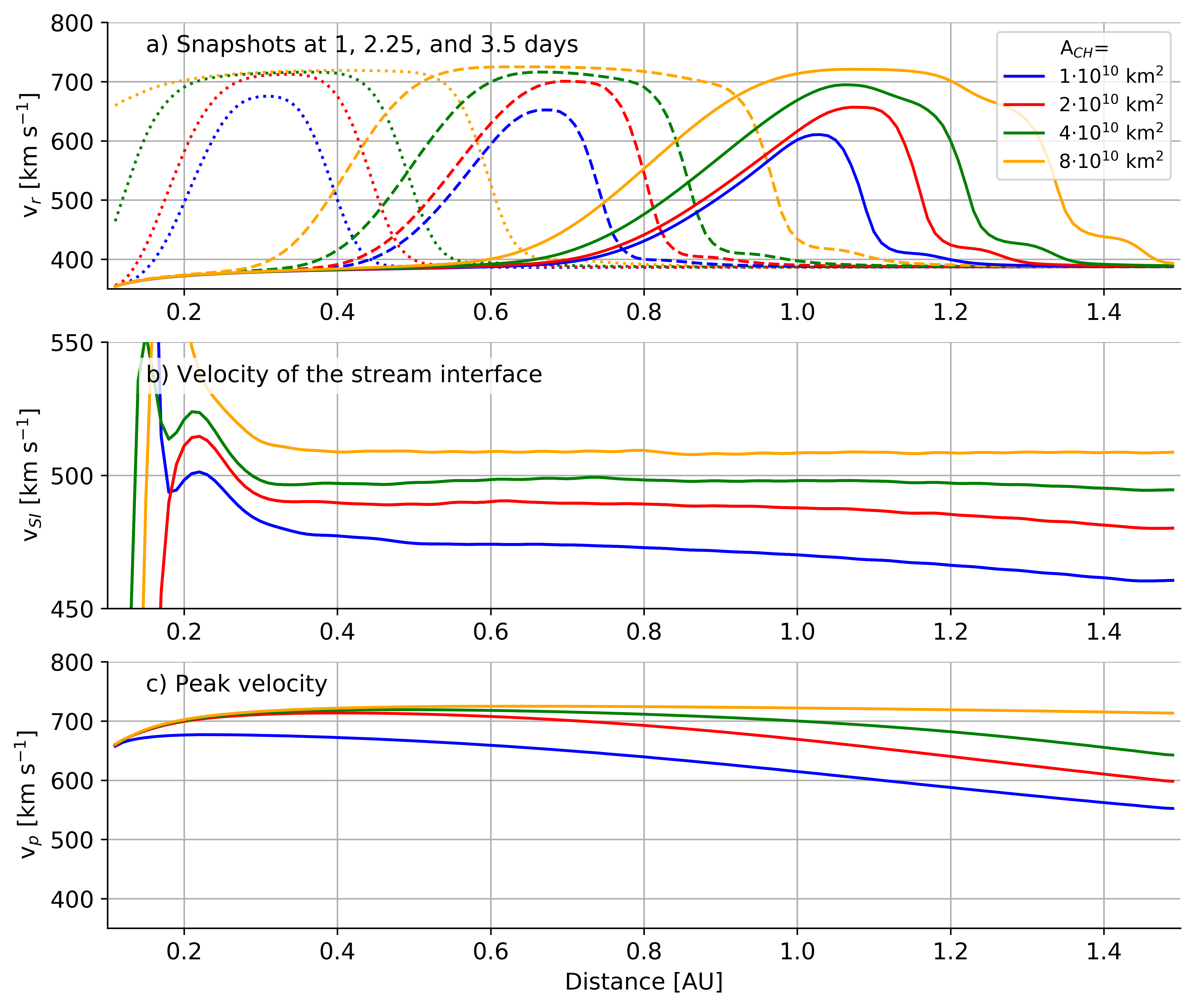}
    \caption{Velocities in the latitudinal centers of HSSs, belonging to coronal holes with areas of $1$ (blue), $2$ (red), $4$ (green), and $8\cdot$\SI{e10}{km^2} (yellow). a) Snapshots of the radial velocity profile, taken at $1$ day (dotted), $2.25$ days (dashed), and $3.5$ days (solid lines) after the center of the coronal hole has crossed the central meridian. b) Velocity of the stream interface $v_\text{SI}$ derived by its radial displacement with time versus the distance of the stream interface to the Sun. c) Peak velocity $v_p$ of the HSS versus the radial distance of the location of the peak velocity to the Sun.}
    \label{fig:radialevo_center}
\end{figure*}

\begin{figure*}[t]
    \centering
   \includegraphics[width=\textwidth]{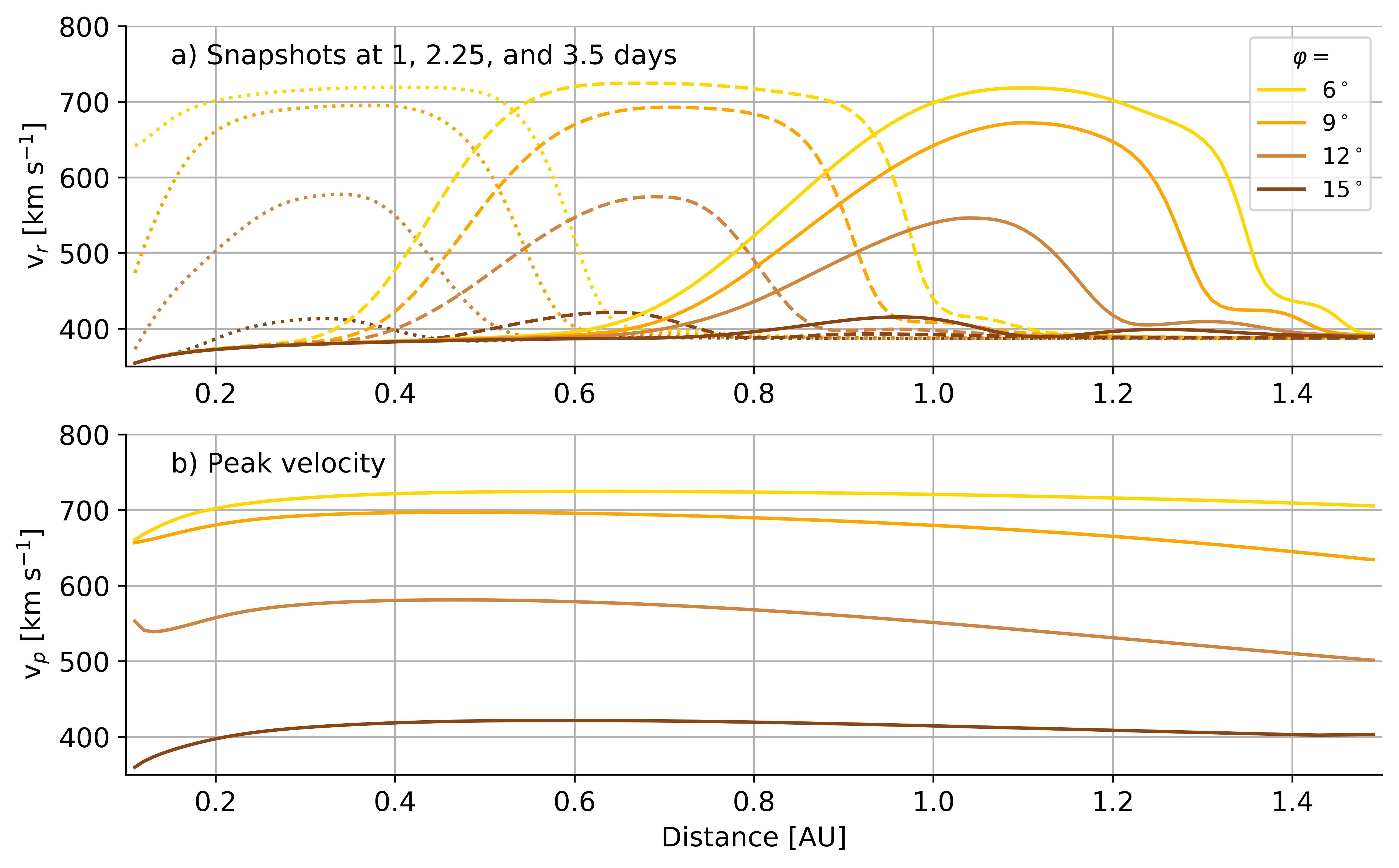}
    \caption{Velocities in the flank of the HSS belonging to the coronal hole with an area of \SI{8e10}{km^2}. The velocities were determined at latitudinal displacements of \SI{6}{\degree} (yellow), \SI{9}{\degree} (orange), \SI{12}{\degree} (brown), and \SI{15}{\degree} (dark brown) to its latitudinal center. a) Snapshots of the radial velocity profiles in the the flank, taken at $1$ day (dotted), $2.25$ days (dashed), and $3.5$ days (solid lines) after the center of the coronal hole has crossed the central meridian. b) Peak velocities $v_\text{p}$ of the HSS in the flank versus the radial distance of the location of the peak velocity to the Sun.}
    \label{fig:radialevo_flanks}
\end{figure*}

In the previous sections, we studied how the HSS peak velocities measured in the solar equatorial plane at \SI{1}{AU} depend on the area and latitude of the source coronal hole. Here, we investigate where these relationships form.
In Figure \ref{fig:radialevo_center} a, we show three snapshots of the velocity profile in the center of HSSs along a radial direction, taken $1$ day, $2.25$ days, and $3.5$ days after the center of the coronal hole has crossed the central meridian. It is apparent that the shape of the velocity profiles get skewed toward the out-flowing direction, which is a result of the dispersion of the flows with radial distance. Furthermore, the peak velocity, in particular of the fastest stream belonging to the largest coronal hole of \SI{8e10}{km^2}, stays almost constant. 

We note that the velocity of propagation of the overall HSS is not given by its peak velocity, but by the velocity of the stream interface between the HSS and the preceding slow solar wind plasma, which is shown in Figure \ref{fig:radialevo_center} b. Since the peak velocities of HSSs are faster than the velocity of the stream interface, the fastest plasma regions in the center of HSSs ultimately collide and are decelerated by the stream interaction region. Then, the new peak velocity is given by the fastest plasma regions which did not interact with the stream interaction region yet. For very extended HSS, which show an extended plateau in the radial velocity profile, this results in only the leading region of the plateau becoming decelerated. Consequently, the position of the peak velocity in the HSS shifts from the front further to its back, while the peak velocity itself stays roughly constant (Fig. \ref{fig:radialevo_center} a and c, yellow lines). For more usual HSSs not having an extended plateau, this results in a continuous impingement of the fastest plasma cells into the stream interaction region, and consequently a decrease of the peak velocity with increasing distance to the Sun (Fig. \ref{fig:radialevo_center} c, red, blue, and green lines). 

 Finally, in Figure \ref{fig:radialevo_flanks}, we show the radial evolution of the peak velocities in the flanks of the HSS that result from the large coronal hole of $A_\text{CH} = \SI{8e10}{km^2}$. The peak velocities are derived at latitudinal displacements of \SI{6}{\degree}, \SI{9}{\degree}, \SI{12}{\degree}, and \SI{15}{\degree} to the center of the HSS. Besides the decrease of peak velocities with radial distance, it is apparent that already very close to the Sun the peak velocities in the flanks of the HSS are much smaller. This is in particular noteworthy since we have set the initial velocity of the HSS homogeneously to \SI{650}{km.s^{-1}}. Therefore, the initial chosen velocity step function at the boundary between the HSS and the ambient slow solar wind erodes very quickly close to the Sun, forming the latitudinal velocity profile of the HSS.

\section{Discussion and Conclusions} \label{sec:conclusions}
In this study, we have simulated the propagation of HSSs in the inner heliosphere, and compared their properties to the empirical relationships between the coronal hole sizes and HSS peak velocities known from observations. To this aim, we used an idealized simulation setup and parametric study, where we have projected the areas of coronal holes to a sphere at \SI{21.5}{\rsun} defining the cross-sections of the associated HSSs, set the plasma and magnetic properties of all resulting HSSs in the same way, and propagated these streams to \SI{1}{AU} using EUHFORIA MHD simulations. We found that:
\begin{itemize}
    \item the peak velocities of the HSSs, measured by the virtual spacecraft in the solar equatorial plane at \SI{1}{AU}, scale linearly with the areas of the source coronal holes for each of the presumed coronal hole latitudes $\varphi = \SI{0}{\degree}, \SI{3}{\degree}, \SI{6}{\degree}, \SI{9}{\degree}, \SI{12}{\degree}$,
    \item the slope of this linear relationship decreases with increasing latitude of the source coronal hole, 
    \item the peak velocities saturate at a constant value of about \SI{730}{km.s^{-1}}.
\end{itemize}
The empirical relationship between the peak velocities of HSSs and the areas of their source coronal holes can therefore be reproduced by our simulations. Since we have neglected the acceleration phase of HSSs and only modelled their propagation from the Sun to Earth, it follows that this empirical relationship describes the propagational evolution of HSSs on their way from the Sun to Earth. Consequently, this empirical relationship is not related to the acceleration of HSSs. 
By looking for the cause of this dependency, we found that:
\begin{itemize}
    \item the maximum velocities of the overall HSSs, determined in their center, depend on the latitudinal width of the source coronal hole up to a width of \SI{10}{\degree} due to erosion effects from the flanks which appear very close to the Sun,
    \item the maximum velocities further depend on the longitudinal width of the coronal hole, which can be explained by the dispersion of the streams during their propagation away from the Sun, and "absorption" of the fastest plasma regions by the stream interaction region.
\end{itemize}

Despite the simplicity of our simulation setup, we can reproduce and explain basic observational relationships between the area and latitude of the source coronal holes on the Sun and the HSS peak velocities in the solar equatorial plane at \SI{1}{AU}. However, we also note that our study lacks a few details. First, for getting the cross-section of the HSS at the inner boundary at \SI{21.5}{\rsun} by simple projection of the coronal hole area, we have assumed a constant expansion factor of 1. However, in general, each coronal hole could have its own expansion factor dependent on its magnetic flux density and the global magnetic field configuration. Second, we have assumed a radial magnetic field at \SI{21.5}{\rsun}, resulting in a radially propagating HSS. In fact, the magnetic field at this height is inclined due to the interaction with the heliospheric current sheet. This results in the HSSs being bent towards the solar equatorial plane, which affects the relative position of the HSS with respect to the measuring satellite. This also means that the very low coronal hole latitudes \SI{<12}{\degree} we assumed, combined with the radially propagating HSSs employed in our study, correspond to real coronal holes located at higher latitudes, in which the resulting HSSs are bent towards the solar equatorial plane.
Concerning these issues, our study stays qualitatively valid as long as we can assume that the expansion factor and the inclination of the magnetic field does not strongly change during the evolution of the individual coronal holes. Then, we still find for each coronal hole individually during its evolution the linear relationship between the HSS peak velocity and the coronal hole area, and only the exact values of the slopes are affected due to a changed relative position of the satellite within the HSS. 
Despite the simplicity of our assumptions, our study gains its justification by the results, being able to reproduce the empirical relationship between the HSS peak velocities and coronal hole areas, and thus to study its physical origin. 

\acknowledgements
We thank Tadgh Garton for providing his data for Figure \ref{fig:obsfacts}.
S.H. thanks the Austrian Agency for International Mobility (OeAD) for their support via a Marietta-Blau-fellowship.
This project has received funding from the European Union’s Horizon 2020 research and innovation programme under grant agreement No 870405. 
These results were also obtained in the framework of the projects C14/19/089  (C1 project Internal Funds KU Leuven), G.0A23.16N (FWO-Vlaanderen), C~90347 (ESA Prodex), Belspo BRAIN project BR/165/A2/CCSOM, and P27292-N20 (Austrian Science Fund). EUHFORIA is developed as a joint effort between the University of Helsinki and KU Leuven. The full validation of the solar wind and CME modelling is being performed within the CCSOM project (http://www.sidc.be/ccsom/). The computational results presented have been achieved using the Vlaams Supercomputer Centrum and the Vienna Scientific Cluster.

\end{document}